\documentclass[twocolumn,aps,showpacs,floatfix,prc]{revtex4}
\usepackage[dvips]{epsfig}
\begin{document}
%\draft
\title{Theoretical exploration of $S$-factors for nuclear reactions of astrophysical importance}

\author{ Vinay Singh$^{1\S\dagger}$, Joydev Lahiri$^{2\S}$ and D. N. Basu$^{3\S\dagger}$}

\affiliation{$^{\S}$Variable Energy Cyclotron Centre, 1/AF Bidhan Nagar, Kolkata 700064, INDIA}
\affiliation{$^\dagger$Homi Bhabha National Institute, Training School Complex, Anushakti Nagar, Mumbai 400085}

\email[E-mail 1: ]{vsingh@vecc.gov.in}
\email[E-mail 2: ]{joy@vecc.gov.in}
\email[E-mail 3: ]{dnb@vecc.gov.in}

\date{\today }

\begin{abstract}

    We present here a robust analytical model based on nuclear reaction theory for non-resonant fusion cross sections near Coulomb barrier. The astrophysical $S$-factors involving stable and neutron rich isotopes of C, O, Ne, Mg and Si for fusion reactions have been calculated in the centre of mass energy range of 2-30 MeV. The model is based on the tunneling through barrier arising out of nuclear, Coulomb and centrifugal potentials. Our formalism predicts precisely the suppression of $S$-factor at sub-barrier energies which are of astrophysical interest. The cross sections can be convoluted with Maxwell-Boltzmann distribution of energies to obtain thermo- or pycno- nuclear reaction rates relevant to nucleosynthesis at high density environments and stellar burning at high temperatures as well as for $^{34}$Ne $+$ $^{34}$Ne fusion occurring in the inner crust of accreting neutron stars.

\vskip 0.2cm


\end{abstract}

\maketitle

\noindent
\section{Introduction}
\label{section1}

    The nuclear reaction cross sections and its convolution with Maxwell-Boltzmann distribution of energies are important for modeling many physical phenomena occurring under extreme conditions \cite{Bu57,Fo64,Cl83}. Such environments of very high temperature or density exist in main-sequence stars and compact stars which are in final stages of their evolutionary development. The exothermic nuclear fusion drives nuclear explosions in the surface layers of the accreting white dwarfs (nova events), in the cores of massive accreting white dwarfs (type Ia supernovae) \cite{Ni97,Ho06} and in the surface layers of accreting neutron stars (type I X-ray bursts and superbursts \cite{St06,Sc03,Cu06,Gu07}). The type-I X-ray bursts and the nova events are generally produced by stellar burning of hydrogen in the thermonuclear regime, without significant effect of plasma screening on the Coulomb tunneling of interacting nuclei. The  superbursts and type Ia supernovae are driven by the burning of C, O, Ne, Si and heavier elements \cite{St06,Sc03,Cu06,Gu07} at high densities, where the plasma screening effect may be significant. In the inner crust of accreting neutron stars (in binaries with low-mass companions \cite{Gu07,Ha90,Ha03}) the pycnonuclear burning of neutron rich nuclei such as $^{34}$Ne $+$ $^{34}$Ne is most likely the source of their internal heat. All these astrophysical processes require precise knowledge of nuclear reaction rates.  

    The thermonuclear reaction rates can be obtained by convoluting fusion cross sections with Maxwell-Boltzmann distribution of energies. These cross sections can vary by several orders of magnitude across the required energy range. The sharp energy dependence of these cross sections can be accounted by an exponential factor while the astrophysical $S$-factor is a smooth slowly varying function of energy facilitating its extrapolation down to astrophysical energies. The low energy fusion cross sections $\sigma$ can only be obtained from laboratory experiments, some of which are not as well known. The theoretical estimates of the thermonuclear reaction rates depend on the various approximations used. Several factors influence the measured values of the cross sections. We need to account for the Maxwellian-averaged thermonuclear reaction rates in the network calculations used in Big-Bang and stellar nucleosynthesis.       
    
    At low energies where the classical turning point is much larger than the nuclear radius, barrier penetrability can be approximated by $\exp(-2\pi\eta)$ so that the charge induced cross section can be decomposed into 

\begin{equation}
 \sigma(E) = \frac{S(E)\exp(-2\pi\eta)}{E}
\label{seqn1}
\end{equation}
\noindent
where $E$ is the centre-of-mass energy, $S(E)$ is the astrophysical $S$-factor and $\eta$ is the Sommerfeld parameter, defined by $\eta = \frac{Z_1Z_2e^2}{\hbar v}$ where $Z_1$ and $Z_2$ are the charges of the interacting nuclei in units of elementary charge $e$. In case of a narrow resonance, the resonant cross section is approximated generally by a Breit-Wigner expression, whereas the neutron induced reaction cross sections at low energies can be given by $\sigma(E)=\frac{R(E)}{v}$ \cite{Bl55} where $R(E)$ is a slowly varying function of energy \cite{Mu10} and is similar to $S$-factor. The astrophysical $S$-factor $S(E)$ which is thus a rescaled variant of the total cross section $\sigma(E)$ is required for many astrophysical applications particularly at energies below the Coulomb barrier. It is much easier to extrapolate $S(E)$ down to low energies of astrophysical interest than the cross section $\sigma(E)$. 

    Depending on the temperature and the density, along with other parameters, stellar burning may involve reactions of different kind from light to heavy nuclei, and from stable to unstable proton or neutron rich. The rates of these reactions can be calculated from the reaction cross sections $\sigma$ by averaging over a Maxwell-Boltzmann distribution of energies. The Maxwellian-averaged thermonuclear reaction rate $<\sigma v>$ at temperature $T$, is given by the following integral \cite{Bo08}:

\begin{equation}
 <\sigma v> = \Big[\frac{8}{\pi\mu (k_B T)^3 } \Big]^{1/2} \int \sigma(E) E \exp(-E/k_B T) dE,
\label{seqn2}
\end{equation}
\noindent
where $v$ is the relative velocity and $\mu$ is the reduced mass of the reacting nuclei. 

    The nuclear fusion reactions at very low energies play the most important role in the nucleosynthesis of light elements in big bang nucleosynthesis as well as nuclear burning inside the stellar core. In the present work, we have calculated $S(E)$ for a number of fusion reactions of astrophysical importance. The theoretical formulation is based on barrier penetration model. The barrier arising out of nuclear and Coulomb potentials is assumed to be of parabolic shape and the centrifugal barrier is added to it. The energy dependence of the cross sections and astrophysical $S$-factors for the fusion reactions involving stable and several neutron rich isotopes of C, O, Ne, Mg and Si covering a wide range of energy from 2 MeV to 30 MeV, below and above the Coulomb barrier, have been calculated using this model.  

\begin{figure}[t]
\vspace{0.0cm}
\eject\centerline{\epsfig{file=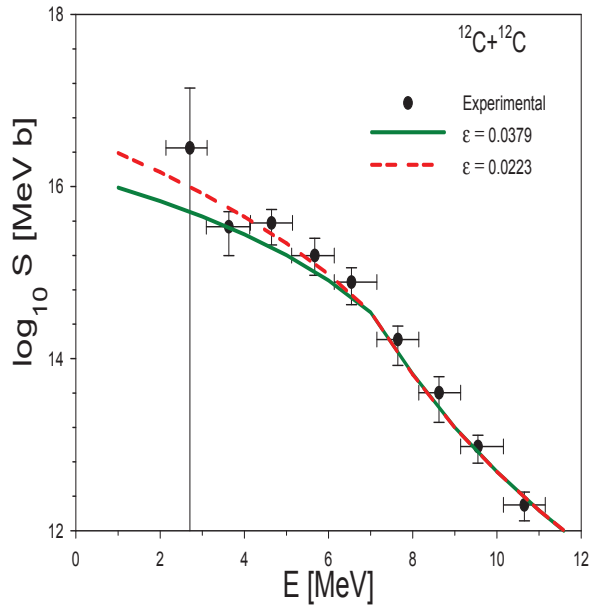,height=8cm,width=7.7cm}}
\caption{(Color online) Plots of $S$-factors for C+C fusion reaction. The filled dots with error bars represent the experimental data points \cite{Pa69,Ma73,Hi77,Ke80,Be81,Er81,Da82,Sa82,Ro03,Ag06,Ba06,Sp07} while the solid and the dashed lines are our present calculations corresponding to two $\varepsilon$ values.}
\label{fig1}
\vspace{0.0cm}
\end{figure}

\begin{figure}[t]
\vspace{0.0cm}
\eject\centerline{\epsfig{file=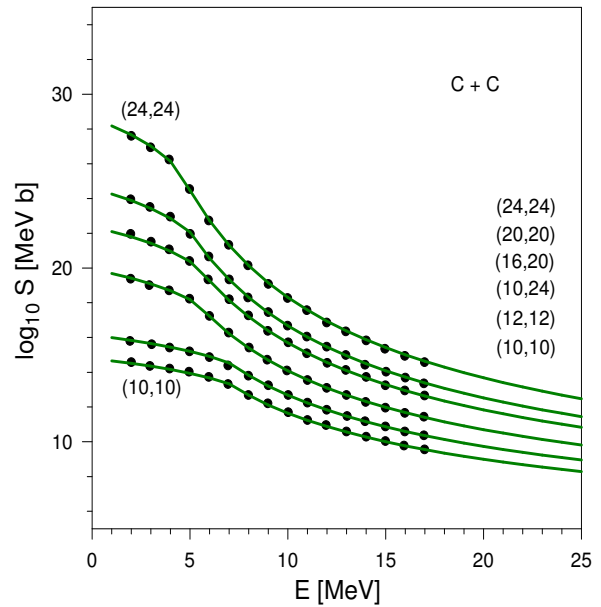,height=8cm,width=7.7cm}}
\caption{(Color online) Plots of $S$-factors for six C+C fusion reactions. The filled dots are the S\~ao Paulo (SP) \cite{ADNDT10} calculations while the solid lines are our present calculations (see Table-I).}
\label{fig2}
\vspace{0.0cm}
\end{figure}

\noindent
\section{Theoretical foundation}
\label{section2}

    The knowledge of astrophysical $S$-factor $S(E)$ for numerous nuclear reactions is the prime requirement for explaining various astrophysical phenomena. The experimental measurements of cross sections $\sigma(E)$ at energies involved are quite often not available due to the fact that the Coulomb barrier suppresses exponentially the cross sections at low energies. The nuclear physics uncertainties of the calculated $S(E)$ can be significant since the theoretical calculations are model dependent. The calculations show that the theoretical estimate of $S(E)$ for a given reaction can vary by many orders of magnitude in the range of energies of astrophysical relevance. Nevertheless for different reactions \cite{Ya06,ADNDT10} also it may as well vary by several orders of magnitude. The primary motivation is to provide a robust theoretical model of $S(E)$ for non-resonant fusion reactions using minimum number of parameters and approximations.

\begin{figure}[t]
\vspace{0.0cm}
\eject\centerline{\epsfig{file=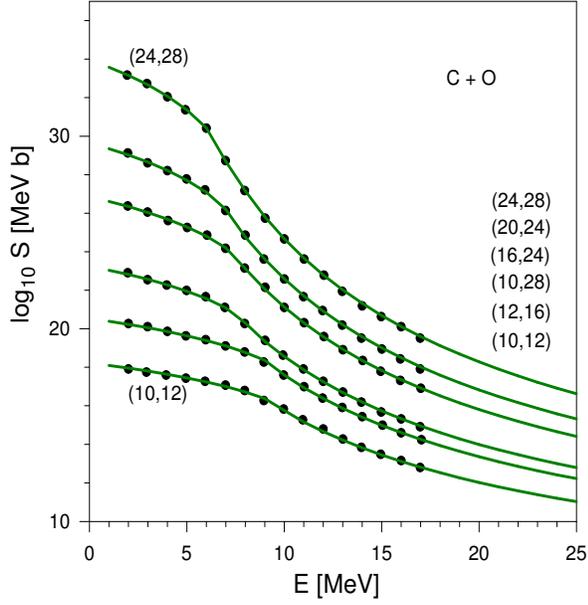,height=8cm,width=7.7cm}}
\caption{(Color online) Plots of $S$-factors for six C+O fusion reactions. The filled dots are the S\~ao Paulo (SP) \cite{ADNDT10} calculations while the solid lines are our present calculations (see Table-I).}
\label{fig3}
\vspace{0.0cm}
\end{figure}

    The present theoretical model is based on the theory of inelastic scattering \cite{La76}. The reaction cross section has been derived accordingly. The transmission coefficient $T_l$ for $l=0$ is calculated assuming quantum mechanical barrier penetration. The potential barrier arising due to nuclear and Coulomb interactions has been assumed to be of a parabolic form. The height and the slope of the barrier are matched at a matching radius $R_m$ greater than the radius $R_{c}$ at which the barrier peaks by an amount characterized by $\varepsilon$. The transmission coefficients for $l>0$ are calculated from physical consideration which accounts for the effect of the centrifugal barrier as well.   

\noindent
\subsection{Ion-ion Nuclear and Coulomb potentials}

    We select an inverse parabolic form for the barrier potential arising due to nuclear and Coulomb interactions. The nuclear force being short range, it is assumed to be an inverse parabolic potential at radial separation $r<R_m$ and pure Coulomb potential beyond:
    
\vspace{0.0cm}
\begin{eqnarray}
\vspace{0.0cm}
V(r)&=&\frac{\lambda}{r}  ~~{\rm for~all}~~r \ge R_m\\ \nonumber
&=& E_c\Big[1-\zeta\frac{(r-R_c)^2}{R_c^2} \Big]~~{\rm for~all}~~r < R_m
\label{seqn3}
\vspace{0.0cm}
\end{eqnarray}
\noindent    
where $\lambda=Z_1Z_2e^2$, $E_c$ is the maximum height of the barrier at $r=R_c$ and $\zeta$ defines the parabolic shape of the barrier. The quantum mechanical requirement demands that the logarithmic derivative of the potential be continuous which implies $V(r)$ and its derivative be continuous at $r=R_m$ which yields 

\vspace{0.0cm}
\begin{eqnarray}
\vspace{0.0cm}
E_c&=&V(R_c)=\frac{\lambda(2+3\varepsilon)}{2R_c(1+\varepsilon)^2},~~~~~\zeta=[\varepsilon(2+3\varepsilon)]^{-1}, \\ \nonumber
E_m&=&V(R_m)=E_c\frac{2(1+\varepsilon)}{2+3\varepsilon} ~~{\rm where}~~\varepsilon=\frac{(R_m-R_c)}{R_c}.
\label{seqn4}
\vspace{0.0cm}
\end{eqnarray}
\noindent
Thus the model is very natural and realistic \cite{Ba02} as well, which allows one to obtain analytic expressions for the barrier penetrability. 

\noindent
\subsection{Quantum tunneling and fusion cross section}
    
    The basic picture of the analytical model is that the fusion cross section can be given by the formal nuclear reaction theory \cite{Bl55}  
    
\begin{equation}
\sigma(E)=\frac{\pi}{k^2} \sum^\infty_{l=0}~(2l+1) T_l(E) f_l(E)
\label{seqn5}
\end{equation}
\noindent
where $k=\frac{\sqrt{2\mu E}}{\hbar}$, $\mu$ being the reduced mass of the interacting nuclei and $T_l(E)$ is the transmission coefficient given by

\begin{equation}
T_l(E)=1-|\eta_l|^2
\label{seqn6}
\end{equation}
\noindent
whereas $f_l(E)$ is the fusion probability of the penetrating wave which at low energies of astrophysical interest is close to unity. The quantity $\eta_l = e^{2i\delta_l}$ where $\delta_l$ is the phase shift for the $l^{th}$ partial wave. In the energy domain involved, the transmission coefficient $T_l(E)$ decreases with $l$ and the largest contribution comes from the $T_0(E)$ term suggesting one to introduce a quantity $N(E)$ such that 

\begin{equation}
N(E)=1+ \sum^\infty_{l=1}~(2l+1)\frac{T_l(E)}{T_0(E)}
\label{seqn7}
\end{equation}
\noindent
whose magnitude is expected to be larger than unity due to the contributions from higher partial waves. 
   
\begin{figure}[t]
\vspace{0.0cm}
\eject\centerline{\epsfig{file=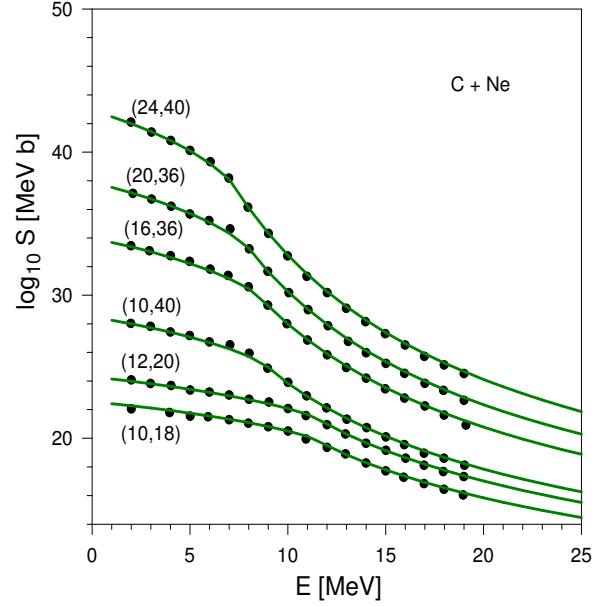,height=8cm,width=7.7cm}}
\caption{(Color online) Plots of $S$-factors for six C+Ne fusion reactions. The filled dots are the S\~ao Paulo (SP) \cite{ADNDT10} calculations while the solid lines are our present calculations (see Table-I).}
\label{fig4}
\vspace{0.0cm}
\end{figure} 

    Substituting Eq.(7) in Eq.(5) one finds that

\begin{equation}
 \sigma(E) = \frac{S_0}{E}N(E)T_0(E)
\label{seqn8}
\end{equation}
\noindent
where $S_0=\frac{\pi\hbar^2}{2\mu}$ and the s-wave transmission coefficient

\begin{figure}[t]
\vspace{0.0cm}
\eject\centerline{\epsfig{file=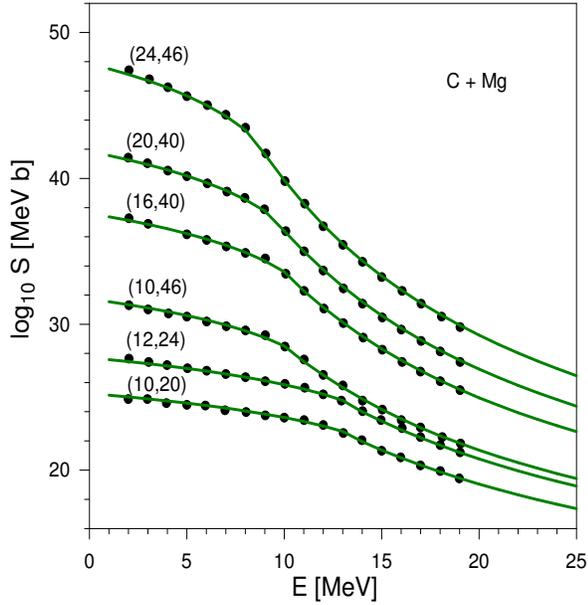,height=8cm,width=7.7cm}}
\caption{(Color online) Plots of $S$-factors for six C+Mg fusion reactions. The filled dots are the S\~ao Paulo (SP) 
\cite{ADNDT10} calculations while the solid lines are our present calculations (see Table-I).}
\label{fig5}
\vspace{0.0cm}
\end{figure}

\vspace{-0.35cm} 
\begin{equation}
T_0(E)=\exp\Big\{-\frac{2}{\hbar}\int^{r_2}_{r_1}\sqrt{2\mu[V(r)-E]}dr\Big\}
\label{seqn9}
\end{equation}
\noindent
where $r_1$ and $r_2$ are classical turning points. Prompted by Eq.(1), expressing $T_0(E)$ as $\exp(\chi-2\pi\eta)$, analytical expressions can be derived for $\chi$ for the barrier potential given by Eq.(3) which for $0 \le E < E_m$ given by

\vspace{-0.35cm}
\begin{eqnarray}
\vspace{0.0cm}
\chi(E)&=&4\sqrt{\frac{E_r}{E}}\Big[\sin^{-1}\sqrt{y_r} + \sqrt{y_r(1-y_r)}\Big]  \\ \nonumber
&-&\xi \sqrt{\frac{E_r}{E_c}}\frac{(E_c-E)}{E_c}\Big[\frac{\pi}{2}+\sin^{-1}y_l+y_l\sqrt{1-y_l^2}\Big]
\label{seqn10}
\vspace{0.0cm}
\end{eqnarray}
\noindent
and for $E_m \le E \le E_{c}$ given by  

\vspace{-0.35cm}
\begin{equation}
\vspace{0.0cm}
\chi(E)=\pi \Big[2\sqrt{\frac{E_r}{E}}-\xi \frac{(E_c-E)}{E_c} \sqrt{\frac{E_r}{E_c}}\Big]
\label{seqn11}
\vspace{0.0cm}
\end{equation}
\noindent
where $y_r=\frac{R_m E}{\lambda}$, $y_l=\varepsilon\sqrt{\frac{\zeta E_c}{E_c-E}}$, $E_r=\frac{\lambda^2\mu}{2\hbar^2}$ and $\xi=\frac{(2+3\varepsilon)^{3/2}\sqrt{\varepsilon}} {(1+\varepsilon)^2}$.   

\begin{figure}[t]
\vspace{0.0cm}
\eject\centerline{\epsfig{file=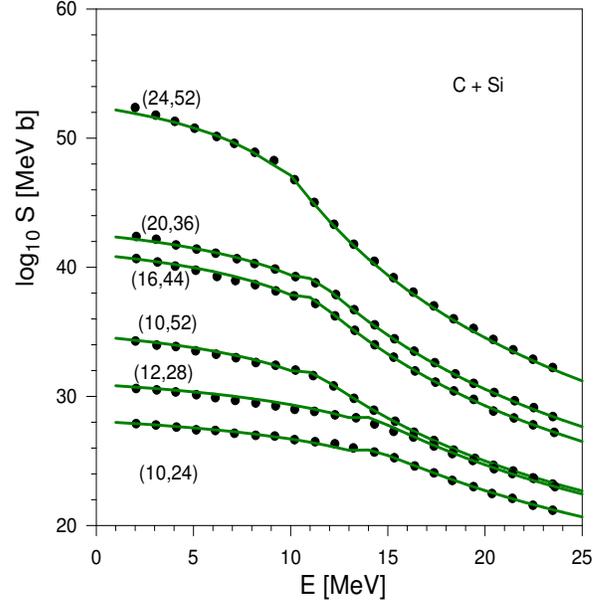,height=8cm,width=7.7cm}}
\caption{(Color online) Plots of $S$-factors for six C+Si fusion reactions. The filled dots are the S\~ao Paulo (SP) \cite{ADNDT10} calculations while the solid lines are our present calculations (see Table-I).}
\label{fig6}
\vspace{0.0cm}
\end{figure}

    The expressions provided above are for sub-barrier energies where the $S$-factor is influenced by several low $l$ values at $E \le E_c$ in which $l=0$ contributing the most. The correcting factor $N(E)$ is expected to be a slowly varying function of energy. It is likely that the transmission coefficients $T_l(E)$ at these $l$ values are similar functions of energy as $T_0$ but of strengths reducing progressively with increasing $l$ implying s-wave like energy dependence. A crude estimate for $N(E)$ at $E \le E_c$ goes as $\sim 1+\sqrt{\frac{E_c}{E_0}}$ \cite{Be12}. To simplify the model it is assumed that $N(E)$ can be approximated by an overall normalization factor  

\begin{equation}
\vspace{0.0cm}
N(E) \approx N_0=1+n_0\sqrt{\frac{E_c}{E_0}}
\label{seqn12}
\vspace{0.0cm}
\end{equation}
\noindent
where $E_0=\frac{\hbar^2}{2\mu R_c^2}$ is the characteristic quantum of centrifugal energy and $n_0$ can be treated as a parameter characterizing the significance of higher partial waves. 

    For above barrier energies $E > E_c$ the effective barrier is transparent for low $l$ waves resulting $T_l=1$. The summation over partial waves in the expression of cross section provided by Eq.(5) goes from $l=0$ to maximum $l=l_m$ at which the effective barrier $V(r)+\frac{\hbar^2l(l+1)}{2\mu r^2}$ becomes classically forbidden. In this case a simplified derivation \cite{Be12} yields

\begin{equation}
\vspace{0.0cm}
\chi(E)= 2\pi \eta + \frac{1}{2} \ln \Big[1 + \frac{y^2(E)}{N_0^2}\Big]
\label{seqn13}
\vspace{0.0cm}
\end{equation}
\noindent    
where $y(E)=\frac{(E-E_c)}{E_0}\Big[1-\frac{(E-E_c)}{\zeta E_c}\Big]$.

    Recalling the definition from Eq.(1), the $S$-factor can be given by

\begin{equation}
\vspace{0.0cm}
 S(E) = N_0S_0\exp\chi(E)
\label{seqn14}
\vspace{0.0cm}
\end{equation}
\noindent  
where the expressions of $\chi(E)$ for different energy domains have been provided by Eq.(10), Eq.(11) and Eq.(13).       
\begin{figure}[ht!]
\vspace{0.0cm}
\eject\centerline{\epsfig{file=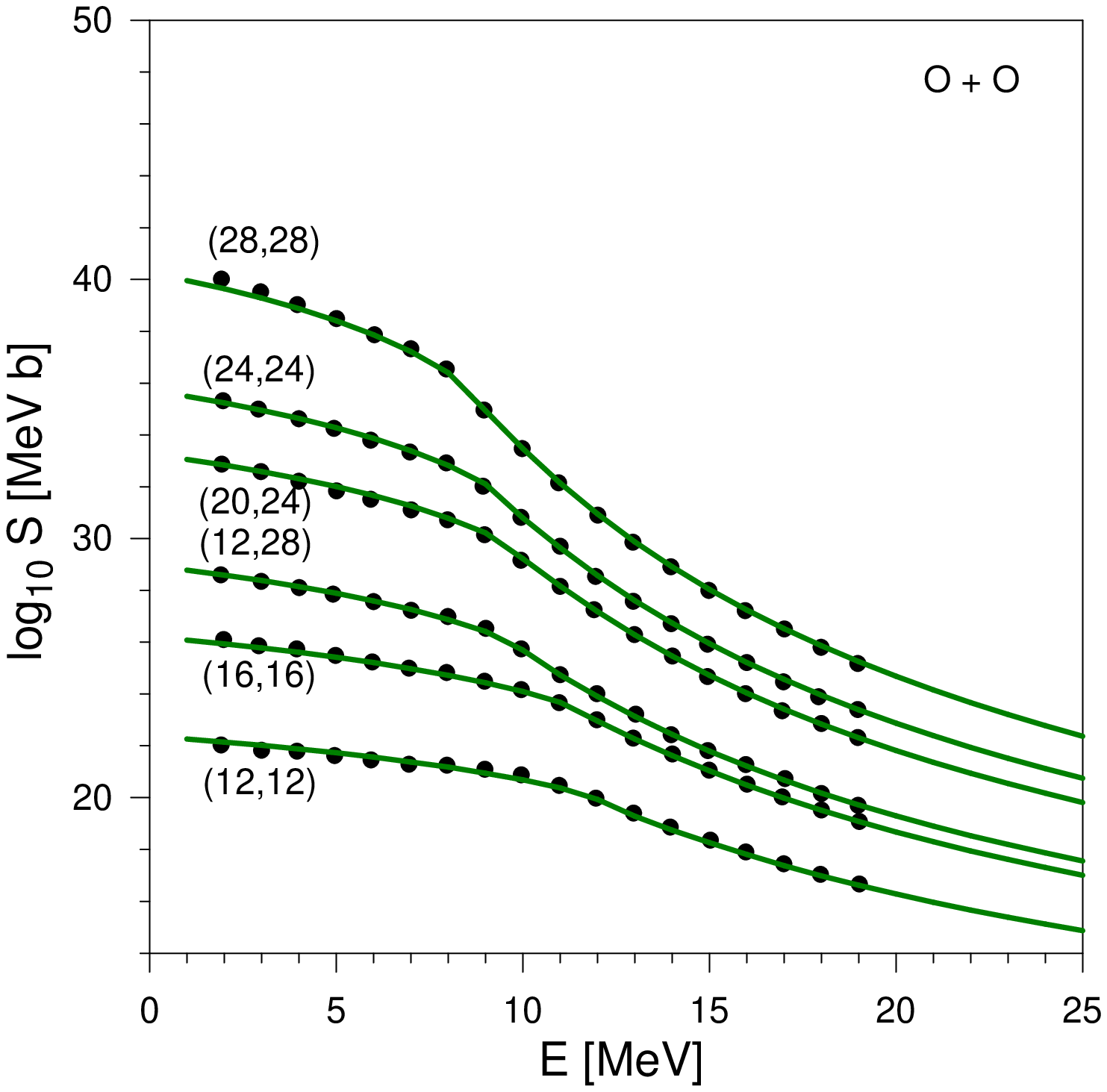,height=7.94cm,width=7.7cm}}
\caption{(Color online) Plots of $S$-factors for six O+O fusion reactions. The filled dots are the S\~ao Paulo (SP) \cite{ADNDT10} calculations while the solid lines are our present calculations (see Table-I).}
\label{fig7}
\vspace{1.0cm}
\end{figure}

\begin{figure}[hb!]
\vspace{0.0cm}
\eject\centerline{\epsfig{file=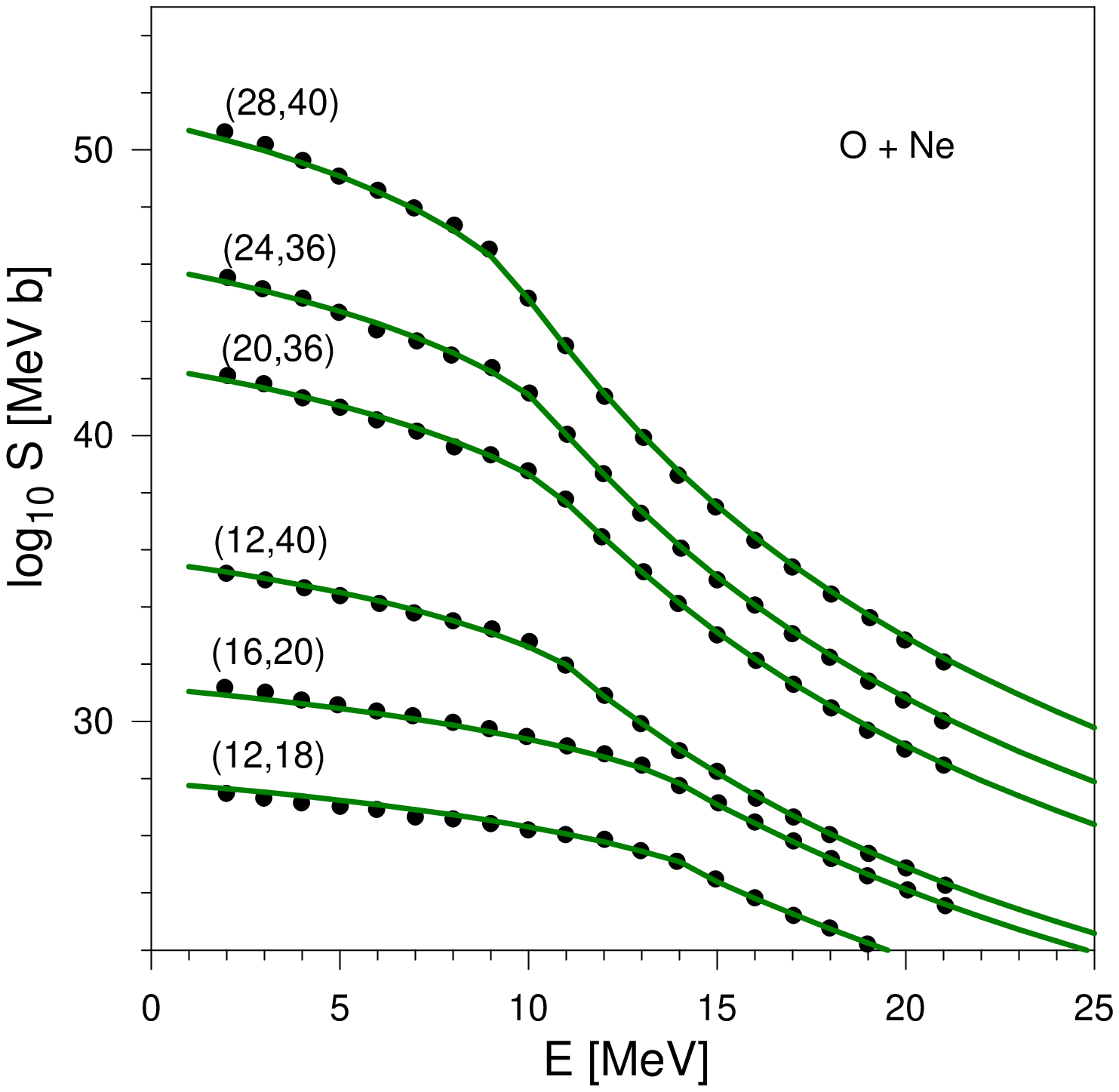,height=7.94cm,width=7.7cm}}
\caption{(Color online) Plots of $S$-factors for six O+Ne fusion reactions. The filled dots are the S\~ao Paulo (SP) \cite{ADNDT10} calculations while the solid lines are our present calculations (see Table-I).}
\label{fig8}
\vspace{0.0cm}
\end{figure} 

\begin{figure}[ht!]
\vspace{0.0cm}
\eject\centerline{\epsfig{file=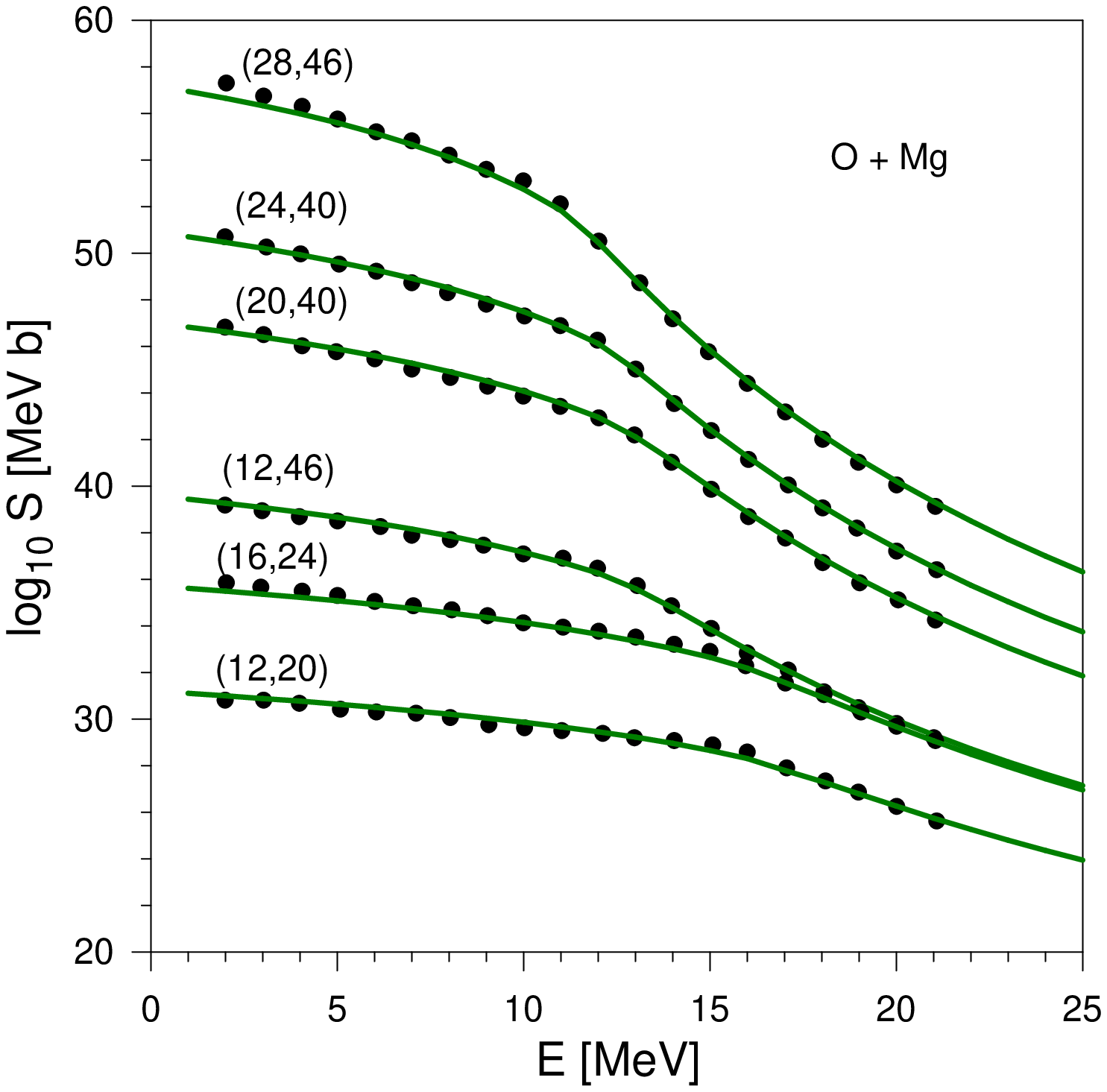,height=7.94cm,width=7.7cm}}
\caption{(Color online) Plots of $S$-factors for six O+Mg fusion reactions. The filled dots are the S\~ao Paulo (SP) \cite{ADNDT10} calculations while the solid lines are our present calculations (see Table-I).}
\label{fig9}
\vspace{1.0cm}
\end{figure} 

\begin{figure}[hb!]
\vspace{0.0cm}
\eject\centerline{\epsfig{file=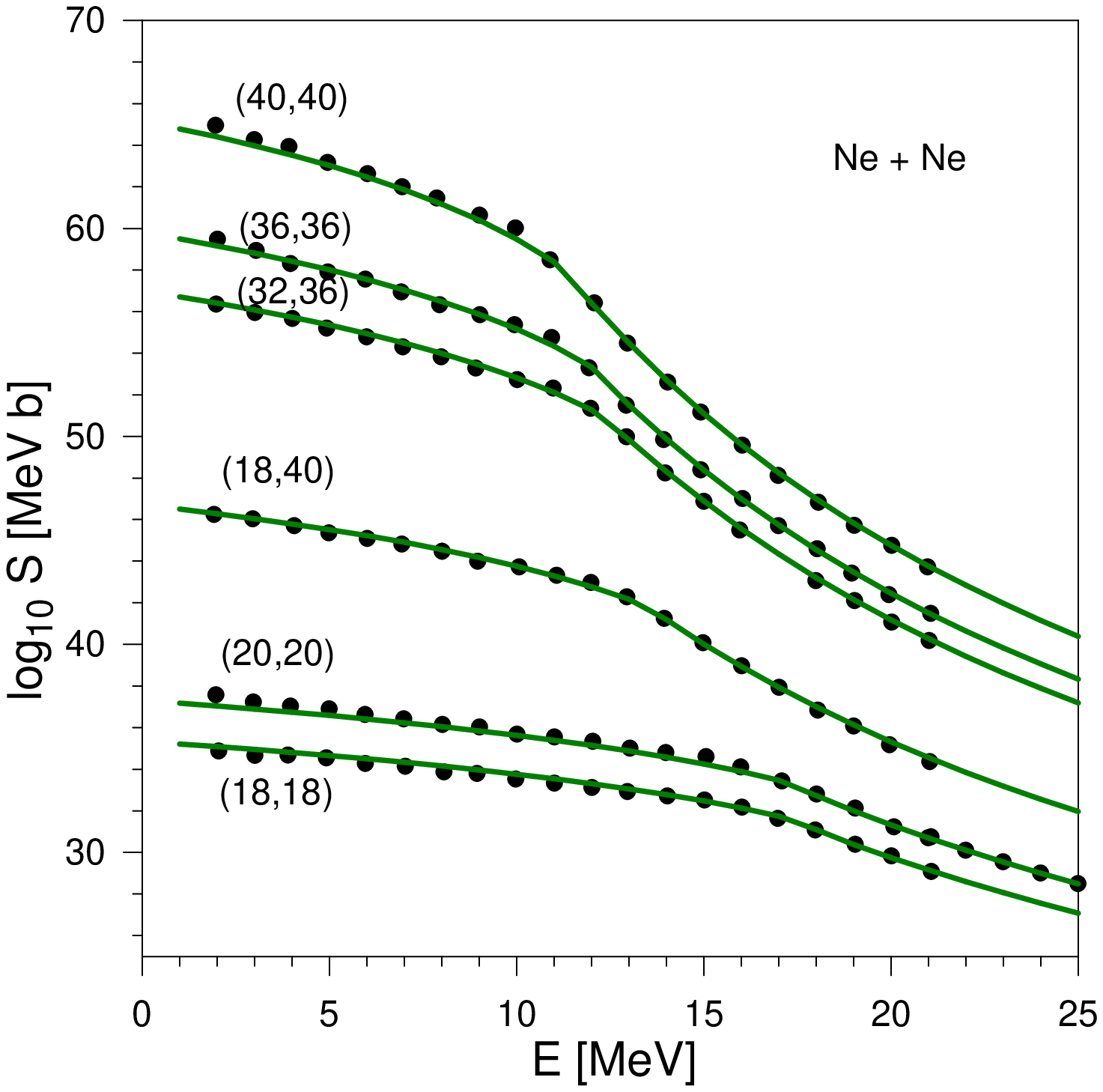,height=7.94cm,width=7.7cm}}
\caption{(Color online) Plots of $S$-factors for six Ne+Ne fusion reactions. The filled dots are the S\~ao Paulo (SP) \cite{ADNDT10} calculations while the solid lines are our present calculations (see Table-I).}
\label{fig10}
\vspace{0.0cm}
\end{figure}  

\begin{figure}[t]
\vspace{0.0cm}
\eject\centerline{\epsfig{file=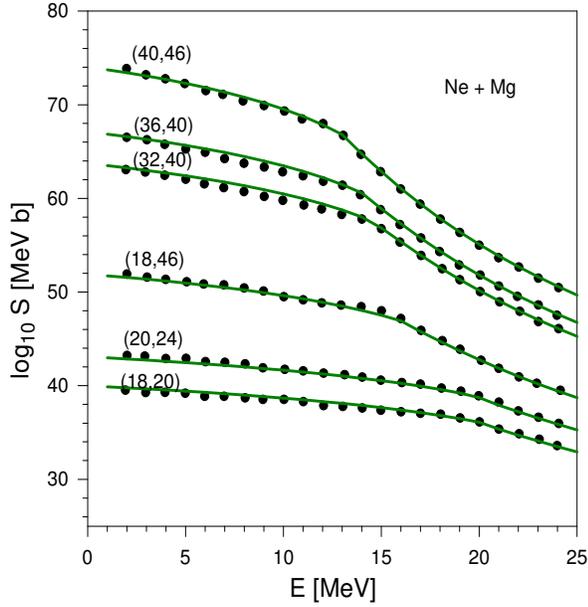,height=8cm,width=7.7cm}}
\caption{(Color online) Plots of $S$-factors for six Ne+Mg fusion reactions. The filled dots are the S\~ao Paulo (SP) \cite{ADNDT10} calculations while the solid lines are our present calculations (see Table-I).}
\label{fig11}
\vspace{0.0cm}
\end{figure}
             
\noindent
\section{Calculation of astrophysical $S$-factor}
\label{section3}

    The calculation of astrophysical $S$-factor using present formalism involves five parameters. The barrier potential is defined by $R_c$ and $\zeta$ (or equivalently $\varepsilon$) since barrier height $E_c$ can be expressed in terms of $R_c$ and $\varepsilon$. The radius at which the barrier peaks can be given in terms of the mass numbers $A_1$ and $A_2$ of the interacting nuclei by $R_c = r_0(A_1^{1/3} +A_2^{1/3}) +|A_1-2Z_1|\Delta_1 +|A_2-2Z_2|\Delta_2$ which involves three parameters. The isotopic dependence of $R_c$ is simulated by entities $\Delta_{1,2}$ and $r_0$ is the radius parameter. The fifth parameter $n_0$ characterizes the effect of partial waves $l>0$. The present formalism causes two fold simplifications: it reduces two parameters and relies on exact theoretical expressions for barrier penetration rather than the approximated ones \cite{Be10,Be12}. The five parameters for various fusing systems have been obtained by fitting the astrophysical $S$-factors from experimental measurements and theoretical results of a nine-parameter phenomenological analytic expression \cite{ADNDT10} which were also compared previously \cite{Ya06,Ga05,Ga07} with experimental data wherever available. The errors in the fitted parameters are calculated from the correlation matrix in the final stage of the fitting procedure when changes in the fitted parameters by amounts equal to the corresponding uncertainties in the fitted parameters cause changes in the corresponding quantity by less than the stipulated value. Thus large uncertainty in a fitted parameter implies that the hyper-surface is rather flat with respect to that parameter. 
 
\begin{figure}[t]
\vspace{0.0cm}
\eject\centerline{\epsfig{file=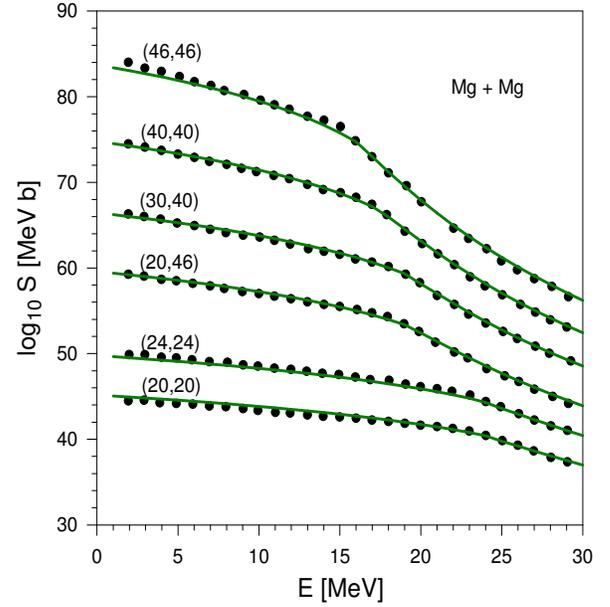,height=8cm,width=7.7cm}}
\caption{(Color online) Plots of $S$-factors for six Mg+Mg fusion reactions. The filled dots are the S\~ao Paulo (SP) \cite{ADNDT10} calculations while the solid lines are our present calculations (see Table-I).}
\label{fig12}
\vspace{0.0cm}
\end{figure}

\begin{table}[htbp]
\vspace{0.0cm}
\centering
\caption{\label{tab:table1} Values of parameters for S(E) for various isotopes of reactions }
\begin{tabular}{|l|c|c|c|c|c|}
\hline
\hline
Reactions &~$r_0$ (fm)~&~~~~~$\varepsilon$~~~~~&~~~~$n_0$~~~~&$\Delta_1$ (fm) &$\Delta_2$ (fm) \\ \hline
 
C+C  	&1.5223 	&0.0379 	&1.7982   &0.1006  	&0.1006 \\
C+O  	&1.5346 	&0.0360 	&1.9156   &0.0967 	&0.0549 \\
C+Ne 	&1.4857 	&0.0326 	&2.3993   &0.1150 	&0.0548 \\
C+Mg 	&1.5162 	&0.0401 	&1.7216   &0.0915 	&0.0524 \\
C+Si 	&1.6732 	&0.0531 	&0.0975   &0.0602 	&0.0481 \\
O+O  	&1.5766 	&0.0410 	&1.5939   &0.0527 	&0.0527 \\
O+Ne 	&1.5628 	&0.0423 	&1.3828   &0.0606 	&0.0555 \\
O+Mg 	&1.5712 	&0.0450 	&0.7472   &0.0481 	&0.0491 \\
Ne+Ne &1.5036	  &0.0387 	&2.1797   &0.0585 	&0.0585 \\
Ne+Mg &1.5237	  &0.0443 	&1.7510   &0.0720 	&0.0399 \\
Mg+Mg &1.4791	  &0.0387 	&1.2801   &0.0473 	&0.0473 \\
 
\hline
\hline
\end{tabular} 
\vspace{0.0cm}
\end{table}
\noindent     
 
\noindent
\section{Results and discussion}
\label{section4}

    The astrophysical $S$-factors involving stable and neutron rich isotopes of C, O, Ne, Mg and Si for fusion reactions have been calculated in the centre of mass energy range of 2-30 MeV. In Fig.-1 the $S$-factors for $^{12}$C+$^{12}$C fusion reaction have been plotted as a function of centre of mass energy of the colliding nuclei. The filled dots with error bars represent the experimental data points \cite{Pa69,Ma73,Hi77,Ke80,Be81,Er81,Da82, Sa82,Ro03,Ag06,Ba06,Sp07}. The solid line represents our present calculations fitted to the S\~ao Paulo (SP) \cite{ADNDT10} calculations. The dashed line has been fitted to the experimental data by just varying the value of $\varepsilon$ while keeping the other four parameters identical as provided in Table-I for C+C system. This exercise has been done to show that the parameters extracted by fitting the S\~ao Paulo (SP) \cite{ADNDT10} results are good enough for predictions of astrophysical $S$-factors. In Figs.-2-12, plots of $S$-factors for various isotopes of C, O, Ne, Mg and Si for fusion reactions have been plotted as a function of centre of mass energy. The filled dots are the S\~ao Paulo (SP) \cite{ADNDT10} calculations while the solid lines are our present calculations. The parameter sets for these reactions are listed in Table-I.

    We find that the standard errors in the fitted parameters are minimum for $r_0$ and $\varepsilon$ and maximum for $n_0$. This implies that the results of the calculations are most sensitive to $r_0$ and $\varepsilon$ and least sensitive to $n_0$. The sensitivity of $\Delta_1$ and $\Delta_2$ on $S(E)$ indicates a possible path of finding isotopic dependence of nuclear radius. The results of the present calculations show that the same set of the five parameters as stated can provide good estimates of $S(E)$ for the entire range of isotopes for a particular combination of interacting nuclei. In case of same nuclei such as C+C or Mg+Mg, the number of parameters further reduces from five to four since in these cases $\Delta_1=\Delta_2$. Moreover, the present formalism not only removes two parameters of Ref.\cite{Be12} but also relies on exact theoretical expressions for barrier penetration rather than the approximated ones used in Refs.\cite{Be10,Be12}.  
          
\noindent
\section{Summary and conclusion}
\label{section5}

    In this work, we present analytical formulation based on barrier penetration model for the astrophysical $S$-factor $S(E)$. The potential barrier due to nuclear and Coulomb interactions has been assumed to be of parabolic nature. The effect of the centrifugal barrier for $l>0$ which in turn means the contributions from higher $l$ values have been simulated phenomenologically. Except for this, the entire formulation is exact and does not invoke any other approximation. Compared to the earlier works \cite{ADNDT10,Be10,Be12}, the present endeavor causes two fold simplifications. It reduces two parameters and relies on exact theoretical expressions for barrier penetration rather than the approximated ones \cite{Be10,Be12}. The energy dependence of the astrophysical S-factors for the fusion reactions involving stable and several neutron rich isotopes of C, O, Ne, Mg and Si covering a wide range of energy from 2 MeV to 30 MeV, below and above the Coulomb barrier, have been calculated. The mentioned reactions are merely a few illustrative examples, but the elegance of the theoretical model in describing the experimental data suggests that it may be used successfully for large number of other nuclei encompassing the entire spectra of isotopes.


\begin{thebibliography}{99}

\bibitem{Bu57} E. M. Burbidge, G. R. Burbidge, W. A. Fowler and F. Hoyle, Rev. Mod. Phys. {\bf 29}, 547 (1957).

\bibitem{Fo64} W. A. Fowler and F. Hoyle, Astrophys. J. Suppl. {\bf 9}, 201 (1964); Appendix C.

\bibitem{Cl83} D. D. Clayton, {\it Principles of Stellar Evolution and Nucleosynthesis} (University of Chicago Press, Chicago, 1983).

\bibitem{Ni97} J. C. Niemeyer, S. E. Woosley, Astrophys. J. {\bf 475}, 740 (1997).

\bibitem{Ho06} P. H\"oflich, Nucl. Phys. {\bf A 777}, 579 (2006).

\bibitem{St06} T. Strohmayer, L. Bildsten, in: W.H.G. Lewin, M. Van der Klis (Eds.), {\it Compact
Stellar X-ray Sources}, (Cambridge University Press, Cambridge, London.), 113 (2006).

\bibitem{Sc03}  H. Schatz, L. Bildsten, A. Cumming, Astrophys. J. {\bf 583}, L87 (2003).

\bibitem{Cu06}  A. Cumming, J. Macbeth, J.J.M. in 't Zand, D. Page, Astrophys. J. {\bf 646}, 429 (2006).

\bibitem{Gu07}  S. Gupta, E.F. Brown, H. Schatz, P. Möller, K.-L. Kratz, Astrophys. J. {\bf 662}, 1188 (2007).

\bibitem{Ha90}  P. Haensel, J.L. Zdunik, Astron. Astrophys. {\bf 229}, 117 (1990).

\bibitem{Ha03}  P. Haensel, J.L. Zdunik, Astron. Astrophys. {\bf 404}, L33 (2003).

\bibitem{Bl55} J. M. Blatt and V. F. Weisskopf, {\it Theoretical Nuclear Physics} (John Wiley $\&$ Sons, New York; Chapman $\&$ Hall Limited, London.) 

\bibitem{Mu10} Tapan Mukhopadhyay, Joydev Lahiri and D. N. Basu, Phys. Rev. {\bf C 82}, 044613 (2010); {\it ibid} Phys. Rev. {\bf C 83}, 067603 (2011).

\bibitem{Bo08} R. N. Boyd, {\it An Introduction to Nuclear Astrophysics} (University of Chicago, Chicago, 2008), 1st ed.

\bibitem{Pa69} J. R. Patterson, H.Winkler and C. S. Zaidins, Astrophys. J. {\bf 157}, 367 (1969).

\bibitem{Ma73} M. G. Mazarakis and W. E. Stephens, Phys. Rev. {\bf C 7}, 1280 (1973).

\bibitem{Hi77} M. D. High and B. Cujec, Nucl. Phys. {\bf A 282}, 181 (1977).

\bibitem{Ke80} K. U. Kettner, H. Lorenz-Wirzba and C. Rolfs, Z. Phys. {\bf A 298}, 65 (1980).

\bibitem{Be81} H. W. Becker, K. U. Kettner, C. Rolfs and H. P. Trautvetter, Z. Phys. {\bf A 303}, 305 (1981).

\bibitem{Er81} K. A. Erb and D. A. Bromley, Phys. Rev. {\bf C 23}, 2781 (1981).

\bibitem{Da82} B. Dasmahapatra, B. Cujec and F. Lahlou, Nucl. Phys. {\bf A 384}, 257 (1982).

\bibitem{Sa82} L. J. Satkowiak, P. A. DeYoung, J. J. Kolata and M. A. Xapsos, Phys. Rev. {\bf C 26}, 2027 (1982).

\bibitem{Ro03} P. Rosales et al., Rev. Mex. F\'is. {\bf 49}, 88 (2003).

\bibitem{Ag06} E. F. Aguilera et al., Phys. Rev. {\bf C 73}, 064601 (2006).

\bibitem{Ba06} L. Barr\'on-Palos et al., Nucl. Phys. {\bf A 779}, 318 (2006).

\bibitem{Sp07} T. Spillane et al., Phys. Rev. Lett. {\bf 98}, 122501 (2007).

\bibitem{Ya06} D. G. Yakovlev, L. R. Gasques, M. Beard, M. Wiescher and A. V. Afanasjev, Phys. Rev. {\bf C 74}, 035803 (2006).

\bibitem{ADNDT10} M. Beard, A. V. Afanasjev, L. C. Chamon, L. R. Gasques, M. Wiescher and D. G. Yakovlev, At. Data Nucl. Data Tables {\bf 96}, 541 (2010).

\bibitem{La76} L. D. Landau and E. M. Lifshitz, {\it Quantum Mechanics}, Pergamon, Oxford, (1976).

\bibitem{Ba02} D. N. Basu, vide figure, e-Print arXiv: nucl-th/0204060 (2002). 

\bibitem{Be12} A. V. Afanasjev, M. Beard, A. I. Chugunov, M. Wiescher and D. G. Yakovlev,  Phys. Rev. {\bf C 85}, 054615 (2012).

\bibitem{Be10} D. G. Yakovlev, M. Beard, L. R. Gasques and M. Wiescher, Phys. Rev. {\bf C 82}, 044609 (2010).

\bibitem{Ga05} L. R. Gasques, A. V. Afanasjev, E. F. Aguilera, M. Beard, L. C. Chamon, P. Ring, M. Wiescher and D. G. Yakovlev, Phys. Rev. {\bf C 72}, 025806 (2005).

\bibitem{Ga07} L. R. Gasques, A. V. Afanasjev, M. Beard, J. Lubian, T. Neff, M. Wiescher and D. G. Yakovlev, Phys. Rev. {\bf C 76}, 045802 (2007).

\end{thebibliography}
\end{document}